\documentclass[amsmath,floatfix,twocolumn,superscriptaddress]{revtex4-1}
\usepackage{amssymb,amsmath}
\usepackage{graphicx,array}
\usepackage{dcolumn}
\usepackage{psfrag}
\usepackage{bm}

\begin{document}

\title{Giant Direct and Inverse Electrocaloric Effects in Multiferroic Thin Films}

\author{Claudio Cazorla}

\affiliation{School of Materials Science and Engineering, UNSW
  Australia, Sydney NSW 2052, Australia} 

\author{Jorge \'I\~niguez}

\affiliation{Materials Research and Technology Department, Luxembourg
  Institute of Science and Technology (LIST), L-4362 Esch/Alzette, Luxembourg}
\affiliation{Physics and Materials Science Research Unit, University of Luxembourg, L-4422 Belvaux, Luxembourg}

\begin{abstract} 
  Refrigeration systems based on compression of greenhouse gases
  are environmentally threatening and cannot be scaled down to on-chip
  dimensions. In the vicinity of a phase transition caloric materials
  present large thermal responses to external fields, which makes them
  promising for developing alternative solid-state cooling devices. Electrocaloric 
  effects are particularly well-suited for portable refrigeration applications; 
  however, most electrocaloric materials operate best at non-ambient 
  temperatures or require the application of large electric fields. Here, we predict 
  that modest electric fields can yield giant room-temperature electrocaloric 
  effects in multiferroic BiCoO$_{3}$ (BCO) thin films. Depending on the orientation 
  of the applied field the resulting electrocaloric effect is either direct (heating) 
  or inverse (cooling), which may enable the design of enhanced 
  refrigeration cycles. We show that spin-phonon couplings and phase competition 
  are the underlying causes of the disclosed caloric phenomena. The dual 
  electrocaloric response of BCO thin films can be effectively tuned by means of 
  epitaxial strain and we anticipate that other control strategies like chemical 
  substitution are also possible. 
\end{abstract}

\maketitle

\section{Introduction}
\label{sec:intro}
Caloric materials react thermally to external fields 
as a result of induced transformations involving sizeable entropy
changes. Their compactness and fast response to external stimuli raise
high hopes to surpass the performance, environmental compliance, and
portability of current refrigeration technologies based on compression
cycles of greenhouse gases. A major breakthrough occurred in the late
nineties with the discovery of giant magnetocaloric effects in
Gd$_{5}$(Si$_{2}$Ge$_{2}$) \cite{pecharsky97}. Yet, most magnetocaloric 
materials are based on scarce rare-earth elements and require
the generation of high magnetic fields. Meanwhile, mechanocaloric
effects are attracting much attention due to their large latent heat
and adiabatic temperature changes \cite{bonnot08,cazorla16,cazorla17a,cazorla17b,cazorla18}. 
The practical implementation of mechanocaloric effects, however, becomes 
difficult when scaling down towards on-chip dimensions. 

In the context of microelectronic cooling, electrocaloric (EC) materials emerge as 
particularly promising owing to their high energy density, natural integration 
in circuitry \cite{scott11,kar-narayan10,defay13,jiang17,geng15,mangeri16,zhang11,tong14,pandya17},
and the possibility of implementing charge-recovery strategies to increase efficiency \cite{defay18}. 
Unfortunately, the largest EC effects observed to date occur at high temperatures 
\cite{mischenko06} or require large electric fields \cite{neese08}, which may lead to 
the appearance of impeding leakage current and dielectric loss problems \cite{defay18,asbani17}. 
Here, we demonstrate giant room-temperature EC effects ($|\Delta T| \sim 10$~K) in epitaxially  
grown BiCoO$_{3}$ (BCO) thin films induced by modest electric fields ($\sim 100$~kVcm$^{-1}$). 
We use first-principles methods \cite{cazorla17,cazorla13} to resolve the phase diagram 
of BCO thin films and show that it is possible to obtain field-driven transitions from 
a non-polar antiferromagnetic (AFM) phase to either a high-entropy polar paramagnetic (PM) 
phase (inverse effect, $\Delta T < 0$) or a low-entropy polar AFM phase (direct effect, 
$\Delta T > 0$), depending on the orientation of the applied electric field. We show that 
the causes of the disclosed EC phenomena are strong phase competition and spin-phonon 
couplings.

\begin{figure*}
\centerline{
\includegraphics[width=1.00\linewidth]{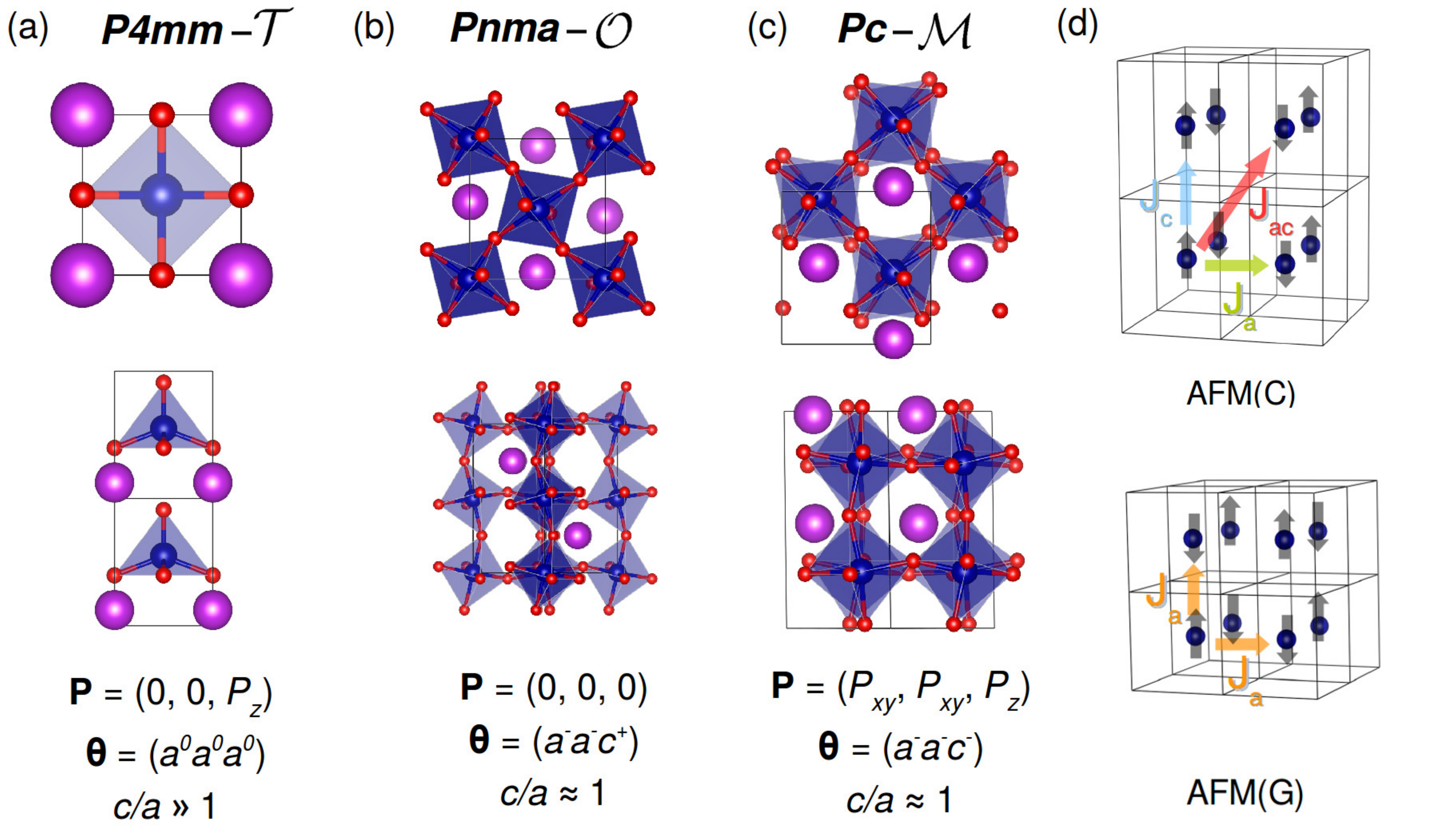}}
\caption{Structural, ferroelectric, and magnetic properties of energetically 
	 competitive bulk BCO polymorphs. (a)~Tetragonal $P4mm$ (${\cal T}$), 
	 (b)~orthorhombic $Pnma$ (${\cal O}$), and (c)~monoclinic $Pc$ 
	 (${\cal M}$). (d)~Sketch of the spin configurations and exchange 
	 constants considered for the ${\cal T}$--AFM(C) and ${\cal O}$--AFM(G)  
	 Heisenberg spin models. Electrical polarizations {\bf P} are referred 
	 to pseudocubic Cartesian axis, and oxygen-octahedra rotation patterns 
	 $\boldsymbol{\theta}$ are expressed in Glazer's notation.}
\label{fig1}
\end{figure*}

\section{Methods}
\label{sec:methods}

\subsection{First-principles calculations}
\label{subsec:fpcal}
In our first-principles calculations, we use the GGA-PBE approximation to density functional theory 
(DFT) \cite{pbe96,vasp} because this functional provides an accurate description of the relative 
stability of bulk BCO polymorphs at zero temperature \cite{cazorla17}. A ``Hubbard-U'' scheme with 
$U = 6$~eV is employed for a better treatment of Co's $3d$ electrons \cite{cazorla17}. We use the 
projector-augmented wave method to represent the ionic cores \cite{bloch94}, considering the 
following electrons as valence states: Co's $3p$, $3d$, and $4s$; Bi's $5d$, $6s$, and $6p$; and 
O's $2s$ and $2p$. Wave functions are represented in a plane-wave basis truncated at $650$~eV and 
for integrations within the first Brillouin zone (BZ) we employ a dense $\Gamma$-centered ${\rm q}$-point 
grid of $12 \times 12 \times 12$ for a $2 \times \sqrt{2} \times \sqrt{2}$ supercell containing 
$20$ atoms (i.e., $4$ formula units) \cite{cazorla15,cazorla17c}. By using these parameters we obtain 
zero-temperature energies converged to within $0.5$~meV per formula unit. Our strained-bulk geometry 
relaxations are performed with a conjugate-gradient algorithm that allows for volume variations while 
imposes the structural constraints defining $(100)$-oriented perovskite thin films \cite{cazorla15,cazorla17c}. 
The relaxations are halted when the forces in the atoms fall all below $0.01$~eV$\cdot$\AA$^{-1}$. The 
calculation of phonon frequencies is performed with the small displacement method \cite{kresse95,alfe09}. 
We find the following settings to provide quasi-harmonic free energies converged to within $5$~meV per 
formula unit: 160--atom supercells, atomic displacements of $0.02$~\AA, and ${\rm q}$-point grids of $16 
\times 16 \times 16$ for BZ integrations. 

\subsection{Estimation of free energies and EC effects}
\label{subsec:free}
We calculate the Helmholtz free energy of competitive polymorphs in $(001)$-oriented BCO thin films 
as a function of temperature ($T$) and in-plane lattice parameter ($a_{\rm in}$) with first-principles 
methods (see works \cite{cazorla17,cazorla13} and Appendix). Our approach 
takes into account the couplings between magnetic spin disorder and lattice phonons, which strongly 
depend on temperature and epitaxial strain. We start by expressing the internal energy of the thin film 
as:
\begin{eqnarray}
\tilde{E}_{\rm harm}(T, a_{\rm in}) & = & \tilde{E}_{0}(T, a_{\rm in}) + \nonumber \\
 & & \frac{1}{2} \sum_{mn} \tilde{\Xi}_{mn}(T, a_{\rm in}) u_{m} u_{n}~,
\label{eq1}
\end{eqnarray}
where $\tilde{E}_{0}$ represents an effective static energy, $\tilde{\Xi}_{mn}$ an 
effective force-constant matrix (see Appendix), $u$'s atomic displacements, 
and the dependences of the various terms on $T$ and $a_{\rm in}$ are explicitly noted. 
The Helmoltz free energy associated to the lattice vibrations, $\tilde{F}_{\rm vib}$, 
is calculated by finding the phonon eigenfrequencies of the dynamical matrix obtained 
from $\tilde{\Xi}_{mn}$, namely, $\tilde{\omega}_{{\bf q}s}$, and subsequently using the 
formula \cite{cazorla17c,cazorla17d}:
\begin{equation}
\tilde{F}_{\rm vib} (T, a_{\rm in}) = \frac{1}{N_{q}}~k_{B} T \sum_{{\bf q}s}\ln\left[ 2\sinh \left( \frac{\hbar \tilde{\omega}_{{\bf q}s}}{2k_{\rm B}T} \right) \right]~,
\label{eq2}
\end{equation}
where $N_{q}$ is the total number of wave vectors used for integration in the BZ, and the summation 
runs over all wave vectors ${\bf q}$ and phonon branches $s$.
The magnetic free energy stemming exclusively from spin fluctuations is estimated through 
the analytical mean-field solution to the spin $\frac{1}{2}$-Ising model at zero magnetic field \cite{strecka15}, 
which reads:
\begin{eqnarray}
\tilde{F}_{mag} (T, a_{\rm in}) & = & -k_{B} T \left[ \ln{2} + \ln{\cosh{\frac{qJm}{2k_{B} T}}} \right] + \nonumber \\
                        &   & \frac{q}{2}Jm^{2}~, 
\label{eq3}
\end{eqnarray}
where $q$ represents the number of nearest-neighboring spins, $J\left(a_{\rm in}\right)$ the exchange 
constant between nearest-neighboring spins, and $m \left(T, a_{\rm in}\right)$ the magnetization of the 
system. In practice, we estimate the value of the magnetic exchange constants from zero-temperature DFT 
calculations (see Appendix) and the magnetization of the system through DFT-based spin-model 
Monte Carlo simulations (see next section). The total Helmholtz free energy of the system then is estimated 
as:
\begin{eqnarray}
\tilde{F}_{\rm harm} (T, a_{\rm in}) & = & \tilde{E}_{0} (T, a_{\rm in}) + \nonumber \\
       & & \tilde{F}_{\rm vib} (T, a_{\rm in}) + \tilde{F}_{mag} (T, a_{\rm in})~.
\label{eq4}
\end{eqnarray} 
Temperature--induced phase transitions are determined via the condition 
$\tilde{F}_{\rm harm}^{A} (T_{c}, a_{\rm in}) = \tilde{F}_{\rm harm}^{B} (T_{c}, a_{\rm in})$,
where $A$ and $B$ represent two different phases. 

The isothermal entropy change that a polar material undergoes under the action of a varying external 
electric field is \cite{moya14}:
\begin{equation}
\Delta S (T, a_{\rm in}) = \int_{0}^{\cal E} \left( \frac{dP}{dT}\right)_{{\cal E}'} d{\cal E}'~, 
\label{eq5}
\end{equation}  
where $P$ represents the polarization of the system and ${\cal E}$ the applied electric field. Likewise, 
the corresponding adiabatic temperature change can be estimated as \cite{manosa17}:
\begin{equation}
\Delta T (T, a_{\rm in}) = -\frac{T}{C_{0}} \Delta S(T, a_{\rm in})~, 
\label{eq6}
\end{equation} 
where $C_{0}(T, a_{\rm in})$ is the heat capacity of the system at zero electric field. $\Delta S$ 
and $\Delta T$ become large when the system undergoes a ${\cal E}$--induced phase transition; if such a 
phase transition is of first-order type, as it occurs in BCO \cite{belik06,oka10}, the isothermal entropy 
change can be estimated with the Clausius-Clapeyron method as \cite{moya14}:
\begin{equation}
\Delta S (T, a_{\rm in}) = -\Delta P \frac{d{\cal E}_{c}}{dT}~, 
\label{eq7}
\end{equation}
where $\Delta P (T, a_{\rm in})$ is the change in polarization along the electric field 
direction, and ${\cal E}_{c} (T, a_{\rm in})$ the critical electric field inducing the phase transformation.  

In the presence of an electric field the thermodynamic potential that appropriately describes the stability 
of a particular phase is the Gibbs free energy, defined as $\tilde{G}_{\rm harm} = \tilde{F}_{\rm harm} - \bm{{\cal E}} 
\cdot \bm{P}$, where $\tilde{F}_{\rm harm}$ corresponds to the Helmholtz free energy in Eq.(\ref{eq4}). In this 
case, the thermodynamic condition that determines a ${\cal E}$--induced phase transition is $\tilde{G}_{\rm harm}^{A} 
(T, a_{\rm in},{\cal E}_{c}) = \tilde{G}_{\rm harm}^{B} (T, a_{\rm in}, {\cal E}_{c})$. The value of the 
corresponding critical electric field then is estimated as:
\begin{equation}
{\cal E}_{c} (T, a_{\rm in}) = \frac{\Delta \tilde{F}_{\rm harm} (T, a_{\rm in})}{\Delta P (T, a_{\rm in})}~,
\label{eq9}
\end{equation}
where $\Delta \tilde{F}_{\rm harm}$ is the Helmholtz free energy difference between the two phases, 
and $\Delta P$ the resulting change in the electric polarization along the electric field direction.  
By using the formulas~(\ref{eq6})--(\ref{eq9}) and knowing $\Delta \tilde{F}_{\rm harm}$, one can 
calculate $\Delta S$ and $\Delta T$ as a function of temperature and in-plane lattice parameter. 

\subsection{Spin-model Monte Carlo simulations}
\label{subsec:MC}
To simulate the effects of thermal excitations on the magnetic order of energetically competitive polymorphs 
in $(100)$-oriented BCO thin films, we construct several spin Heisenberg models of the form $\hat{H} = 
\frac{1}{2} \sum_{ij} J^{(0)}_{ij} S_{i}S_{j}$ based on density functional calculations (see Appendix). 
We use those models to perform Monte Carlo (MC) simulations in a periodically-repeated simulation box of $20 
\times 20 \times 20$ spins; thermal averages are computed from runs of $50,000$ MC sweeps after equilibration. 
These simulations allow us to monitor the $T$-dependence of the magnetic order via the computation of the AFM-C 
and AFM-G order parameters (see Fig.\ref{fig1}d), defined as: $S^{\rm C} \equiv \frac{1}{N} \sum_{i} (-1)^{n_{ix}+n_{iy}} S_{iz}$ and 
$S^{\rm G} \equiv \frac{1}{N} \sum_{i} (-1)^{n_{ix}+n_{iy}+n_{iz}} S_{iz}$. Here, $n_{ix}$, $n_{iy}$, and $n_{iz}$ 
are the three integers locating the $i$-th lattice cell, and $N$ is the total number of spins in the simulation 
cell. For the calculation of $S^{\rm C}$ and $S^{\rm G}$ we consider only the $z$ component of the spins since a 
small symmetry-breaking magnetic anisotropy is introduced in the system Hamiltonian to facilitate the numerical 
analysis.

\begin{figure*}
\centerline{
\includegraphics[width=1.00\linewidth]{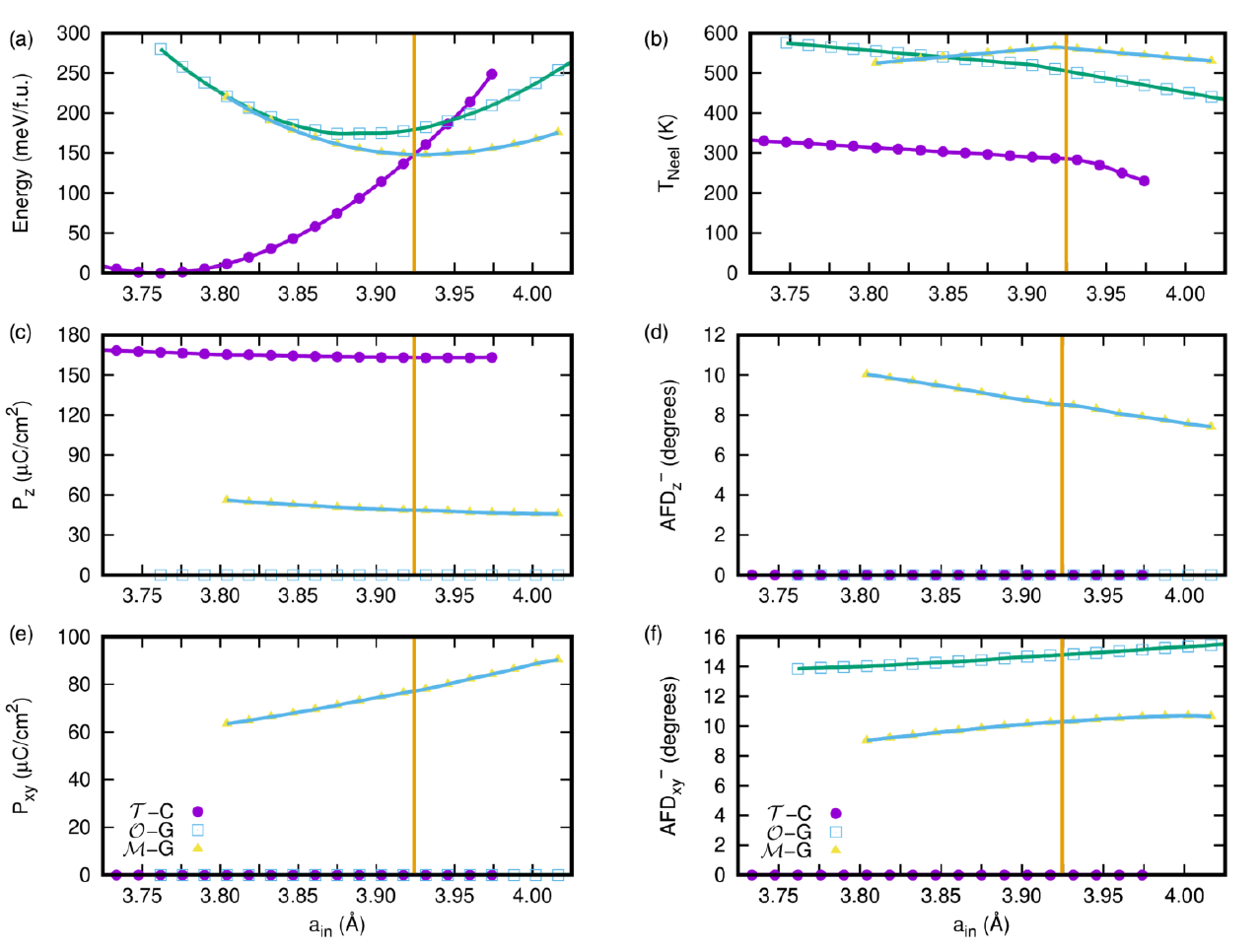}}
\caption{Energy, magnetic, structural, and ferroelectric, properties of
    energetically competitive polymorphs in $(100)$-oriented BiCoO$_{3}$ 
    thin films calculated with first-principles methods.
   (a)~Zero-temperature total energy, (b)~Antiferromagnetic to paramagnetic
   transition temperature, (c)~electric polarization along the $(100)$
   direction, (d)~anti-phase oxygen octahedral rotation angles along the
   $(100)$ direction, (e)~electric polarization along the $(011)$ direction,
   (f)~anti-phase oxygen octahedral rotation angles along the $(011)$ direction. 
   The vertical lines indicate the strain-induced ${\cal T} \to {\cal M}$ 
   phase transition occurring at very low temperatures. Electric polarizations
   are calculated with the Berry phase approach \cite{king93}.}
\label{fig2}
\end{figure*}

\section{Results and Discussion}

\subsection{Phase transitions in BCO thin films}
\label{subsec:pcom}
Figure~\ref{fig1} shows the relevant 
polymorphs of bulk BCO \cite{cazorla17}. At ambient conditions, bulk BCO 
presents the ferroelectric (FE) tetragonal ${\cal T}$ phase shown in
Fig.\ref{fig1}a. This structure has a relatively small in-plane
lattice constant ($c/a \approx 1.3$), hence perovskite substrates with 
small in-plane cell parameters $a_{\rm in}$ will be required for stabilizing 
it in thin-film form. The competing structures are the paraelectric (PE) 
orthorhombic ${\cal O}$ phase (Fig.\ref{fig1}b) and the FE monoclinic 
${\cal M}$ phase (Fig.\ref{fig1}c); both phases have cells that are 
slightly distorted versions of the ideal cubic perovskite structure, 
with $c/a \approx 1$. Note that the mentioned FE phases present spontaneous
polarizations along quite different crystallographic directions --
i.e., pseudocubic $[001]_{\rm pc}$ for ${\cal T}$ and $\sim[111]_{\rm
  pc}$ for ${\cal M}$. As regards magnetism (Fig.\ref{fig1}d), both the
${\cal O}$ and ${\cal M}$ phases display G-type AFM order with a quite 
high N{\'e}el temperature $T_{\rm N} \sim 500$~K. In contrast, the ${\cal T}$ 
phase presents C-type AFM order with a relatively low $T_{\rm N} \sim 310$~K.

\begin{figure*}
\centerline{
\includegraphics[width=1.00\linewidth]{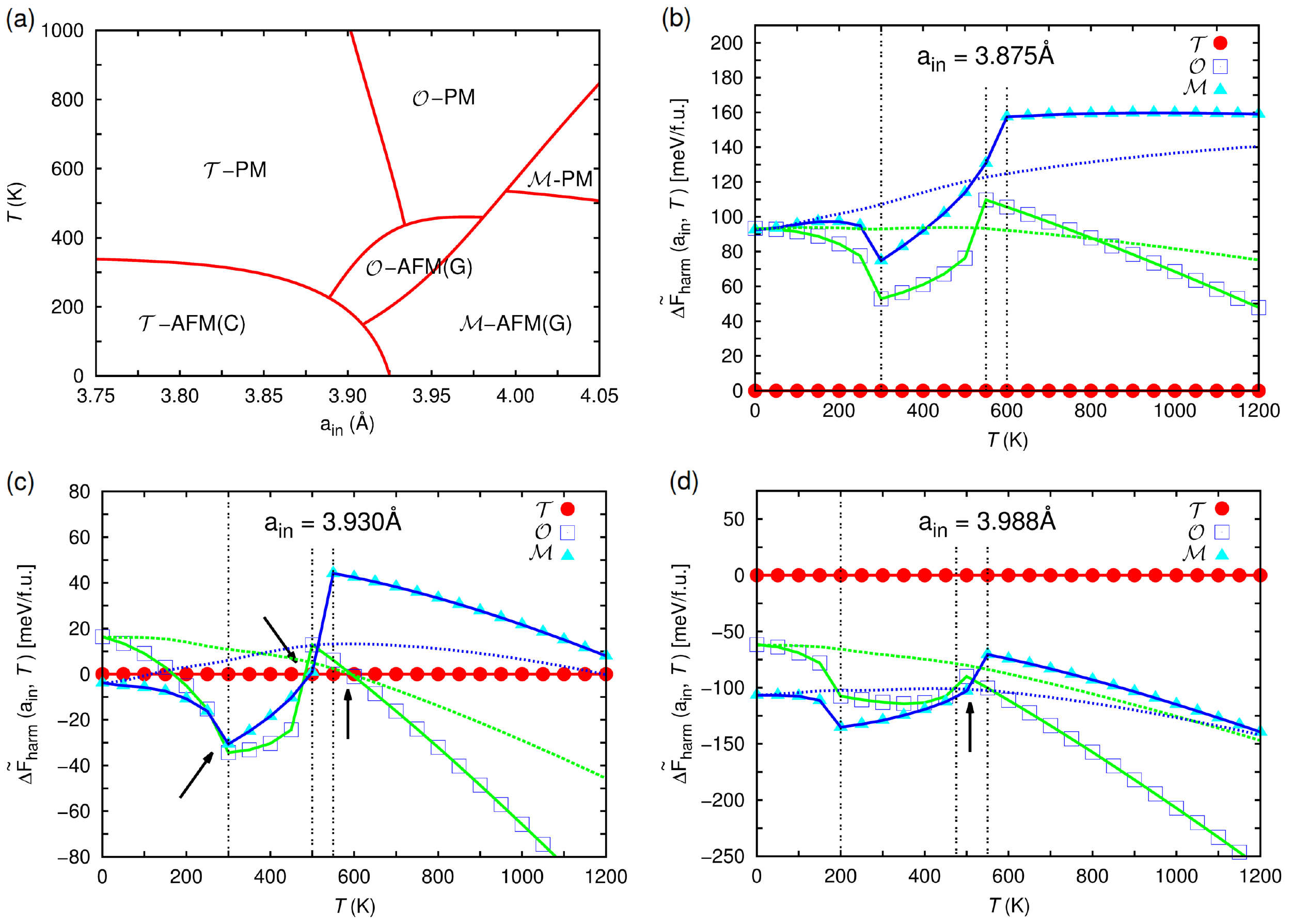}}
\caption{$T$--$a_{\rm in}$ phase diagram and free-energy differences in $(100)$-oriented BCO thin 
  films. (a)~Phase diagram as a function of temperature and in-plane lattice parameter. 
  (b)~Helmholtz free-energy differences among competitive polymorphs at $a_{\rm in} = 3.875$~\AA, 
  (c)~$a_{\rm in} = 3.930$~\AA, (d)~$a_{\rm in} = 3.988$~\AA. 
  Vertical lines and black arrows indicate $T$--induced magnetic and structural phase transitions, 
  respectively. Blue and green dotted lines correspond to Helmholtz free-energy differences 
  for the ${\cal M}$ and ${\cal O}$ phases, respectively, calculated without considering spin-phonon
  coupling effects.}
\label{fig3}
\end{figure*}

We thus expect the competition between BCO polymorphs to be strongly
affected by epitaxially growing thin films on perovskite substrates, as this imposes 
an in-plane lattice constant $a_{\rm in}$ in the system. Our zero-temperature
first-principles calculations confirm this conjecture, yielding the results 
summarized in Fig.\ref{fig2}. We predict a zero-temperature strain-driven ${\cal T} 
\to {\cal M}$ phase transformation at $a_{\rm in} = 3.925$~\AA~ (Fig.\ref{fig2}a) 
that involves rotation of the polarization 
(Figs.\ref{fig2}c,e) and the appearance of anti-phase oxygen octahedral rotations 
along the three pseudocubic directions (Figs.\ref{fig2}d,f). The ${\cal O}$ phase 
remains close in energy to the ${\cal M}$ polymorph over the whole $a_{\rm in}$ 
interval; however, it never becomes stable at zero temperature, in contrast to what 
occurs in compressed bulk BCO \cite{cazorla17}. Meanwhile, the N{\'e}el 
temperature of epitaxially grown BCO thin films decreases mildly with increasing $a_{\rm in}$ 
(Fig.\ref{fig2}b). 

We have performed first-principles Helmholtz free-energy calculations 
of the three relevant BCO polymorphs within the quasi-harmonic approximation 
(see Sec.~\ref{sec:methods} and Appendix), to determine their relative stability
as a function of $T$ and $a_{\rm in}$. Our predicted $T$--$a_{\rm in}$ phase diagram 
for epitaxially grown BCO thin films is shown in Fig.\ref{fig3}a. For relatively small $a_{\rm in}$'s, 
we find that the ${\cal T}$ phase dominates and extends its stability region to temperatures 
much higher than observed in bulk BCO (we recall that bulk BCO presents a ${\cal T} \rightarrow 
{\cal O}$ transition at $T \approx 900$~K \cite{cazorla17}). The reason for this ${\cal T}$ 
stability enhancement is that the competing ${\cal O}$ polymorph remains highly strained at
such $a_{\rm in}$ conditions (Fig.\ref{fig2}a), hence its free energy increases considerably
as compared to the bulk case. 
As $a_{\rm in}$ is increased, the ${\cal T}$ phase eventually is replaced by the ${\cal O}$ and 
${\cal M}$ polymorphs, yielding a very rich phase diagram that exhibits concurrent structural 
and spin-ordering transformations.

Figures~\ref{fig3}b--d show the calculated Helmholtz free-energy differences
$\Delta \tilde{F}_{\rm harm}$ between the three relevant BCO polymorphs expressed
as a function of $T$ and $a_{\rm in}$. As noted previously, our calculations take 
into account all possible sources of entropy, namely, magnetic and vibrational, 
and the interplay between spin disorder and lattice vibrations (see Sec.~\ref{sec:methods} 
and Appendix). We find that at high temperatures ($T \gtrsim 600$~K) the vibrational 
contributions to $\tilde{F}_{\rm harm}$ always favor the ${\cal O}$ and ${\cal M}$ 
phases over the ${\cal T}$ phase. Nonetheless, whenever a polymorph becomes magnetically 
disordered, the corresponding Helmholtz free energy decreases significantly as a consequence 
of $T$--induced lattice phonon softenings (as is appreciated, for instance, in the densities of 
vibrational states calculated at $T < T_{\rm N}$ and $T > T_{\rm N}$, not shown here). 
Accordingly, abrupt $\Delta \tilde{F}_{\rm harm}$ changes appear in Figs.\ref{fig3}b--d 
at the corresponding AFM~$\to$~PM magnetic transition temperatures. The strong spin-phonon 
couplings in epitaxially grown BCO thin films are responsible for the stabilization of the ${\cal O}$--AFM(G) 
phase at temperatures near ambient and $3.88$~\AA~$\le a_{\rm in} \le 3.96$~\AA. Indeed, when spin-phonon 
couplings are disregarded (i.e., quasi-harmonic free energies are calculated in the standard 
manner by considering ``frozen'' spin arrangements) the ${\cal O}$ phase becomes stable 
only after reaching the paramagnetic state at high temperatures (Figs.\ref{fig3}b--d and Fig.\ref{fig2}b).

\subsection{Electrocaloric effects in epitaxially grown BCO thin films}
\label{subsec:EC}
We now focus on the $a_{\rm in}$ region $3.89$~\AA~$\le a_{\rm in} \le 3.93$~\AA, which is 
particularly convenient from a practical perspective since many perovskite substrates 
present lattice constants in this range (where all the phases remain strained by $+4$--$5$\% 
--${\cal T}$-- and $\pm 0.1$--$1$\% --${\cal O}$ and ${\cal M}$-- as compared to their corresponding 
equilibrium in-plane lattice parameters). Interestingly, we find a $T$--driven 
reentrant behavior that is reminiscent of bulk BCO under compression \cite{cazorla17}: upon 
heating, the BCO film transforms first from a FE (${\cal T}$--AFM(C) or ${\cal M}$--AMF(G)) 
phase to a PE (${\cal O}$--AFM(G)) state, then back to a FE (${\cal T}$--PM) phase, and 
finally to a PE (${\cal O}$--AFM(G)) state. Of particular interest is the PE ${\cal O}$--AFM(G) 
region appearing near room temperature $T_{\rm room}$, which is surrounded by two 
FE phase domains presenting markedly different features. In particular, the phase diagram in 
Fig.\ref{fig3}a suggests that the ${\cal O}$ phase may be transformed into the ${\cal T}$ 
or ${\cal M}$ states by applying an electric field ${\cal E}$ along the [001]$_{\rm pc}$ 
or [111]$_{\rm pc}$ directions, respectively. Such ${\cal E}$--driven phase transformations 
involve drastic structural changes as well as magnetic transitions, hence big entropy 
shifts are likely to occur as a consequence.

\begin{figure*}
\centerline{
\includegraphics[width=1.00\linewidth]{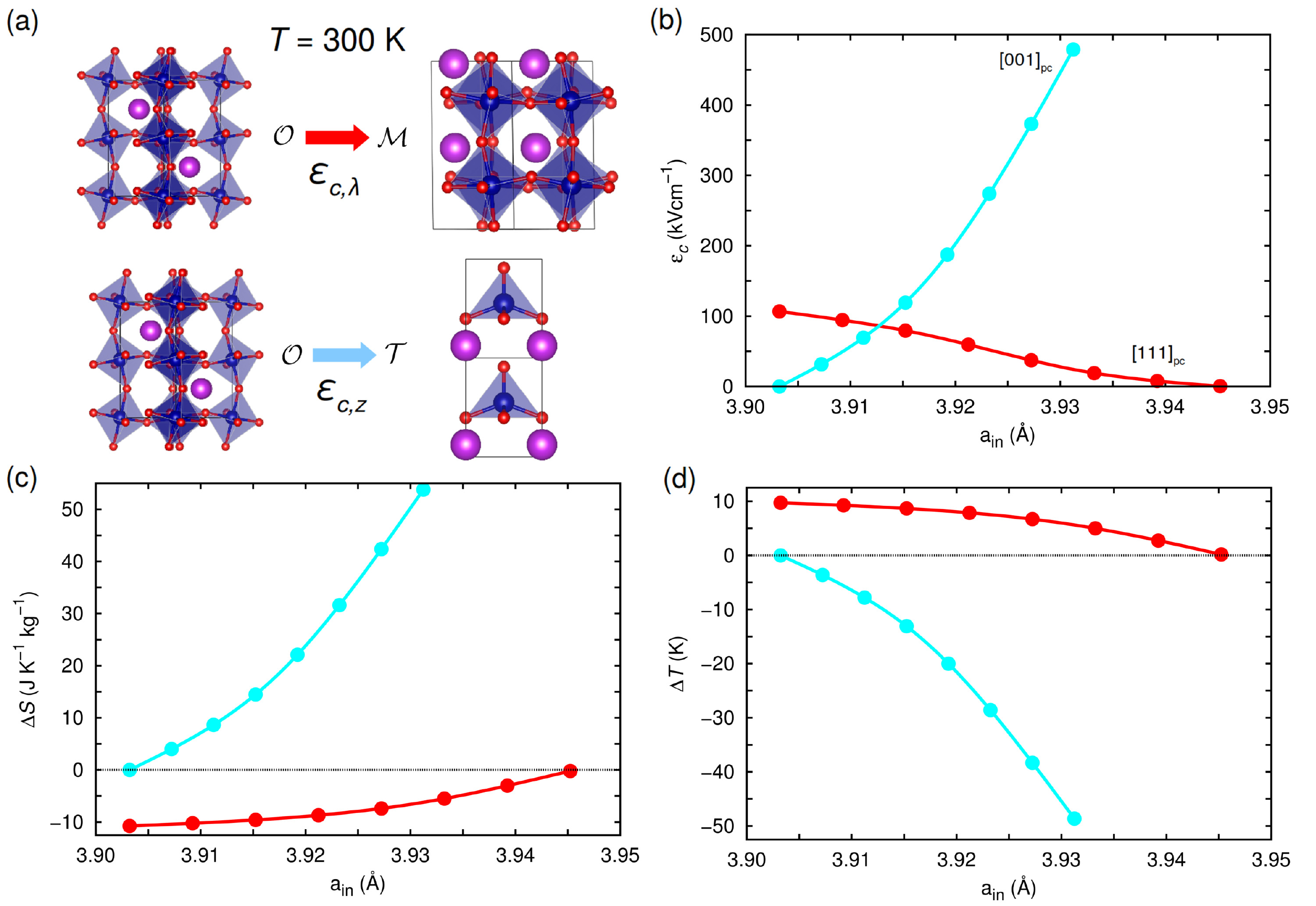}}
\caption{Direct (${\cal O} \to {\cal M}$, red) and inverse (${\cal O} \to {\cal T}$, blue) electrocaloric
         effects in $(100)$-oriented BCO thin films at room temperature. (a)~Sketch of the ${\cal E}$--induced
         phase transformations. (b)~Critical electric field expressed as a function of in-plane lattice parameter;
	 the two involved electric field orientations are indicated in pseudocubic Cartesian notation.
         (c)~Room-temperature entropy and (d)~adiabatic temperature shifts expressed as a function of in-plane
         lattice parameter.}
\label{fig4}
\end{figure*}

Figure~\ref{fig4} shows the direct EC effect associated to the
field-induced ${\cal O} \to {\cal M}$ transformation (Fig.\ref{fig4}a), 
in which the entropy of the system decreases ($\Delta S < 0$, 
Fig.\ref{fig4}c). For the smallest $a_{\rm in}$ values, a maximum adiabatic 
temperature change $\Delta T$ of $+10$~K (Fig.\ref{fig4}d) is estimated for 
a maximum critical electric field of $110$~kVcm$^{-1}$ (Fig.\ref{fig4}b). The 
magnitude of this effect and of the accompanying critical electric field decrease 
with increasing $a_{\rm in}$, as the region of ${\cal M}$ stability is approached. 
Similarly, Fig.\ref{fig4} shows the inverse EC effect associated to the field-induced 
${\cal O} \to {\cal T}$ transformation (Fig.\ref{fig4}a), in which the entropy of 
the film increases ($\Delta S > 0$, Fig.\ref{fig4}c). A maximum $\Delta T$ of $-50$~K 
(Fig.\ref{fig4}d) is estimated for a maximum critical electric field of $500$~kVcm$^{-1}$ 
at $a_{\rm in} = 3.93$~\AA~ (Fig.\ref{fig4}b). The magnitude of this effect and of the
corresponding ${\cal E}_{c}$ decrease with decreasing $a_{\rm in}$, as the region of 
${\cal T}$ stability is approached. We note in passing that application of electric fields 
of the order of $100$~kVcm$^{-1}$ in oxide perovskite thin films appears to be non-problematic 
in practice (see, for instance, Ref.\cite{defay13} and experimental references therein).

\begin{table*}
\centering
\begin{tabular}{c c c c c c c}
\hline
\hline
$ $ & $ $ & $ $ & $ $ & $ $ & $ $ \\
${\rm Material}$ \qquad & \qquad  $T$ \qquad & \qquad $\Delta {\cal  E}$ \qquad & \qquad $C_{0}$ \qquad & \qquad $\Delta S$ \qquad & \qquad $\Delta T$ \qquad & \quad ${\rm Ref.}$ \qquad \\
$ $ \qquad & \qquad ${\rm (K)}$ \qquad & \qquad ${\rm (kVcm^{-1})}$ \qquad & \qquad ${\rm (J K^{-1} Kg^{-1})}$ \qquad & \qquad ${\rm (J K^{-1} Kg^{-1})}$ \qquad & \qquad ${\rm (K)}$ \qquad & $ $ \qquad \\
$ $ & $ $ & $ $ & $ $ & $ $ & $ $ & $ $ \\
\hline
$ $ & $ $ & $ $ & $ $ & $ $ & $ $ & $ $ \\
${\rm PZT}                           $ \qquad & $ 495 $ & $ 776 $ & $ 330 $ & $ -8.0  $ & $ +12.0 $ & \cite{mischenko06}  \\
$ $ & $ $ & $ $ & $ $ & $ $ & $ $ & $ $ \\
${\rm P(VDF-TrFE)}                   $ \qquad & $ 353 $ & $2000 $ & $1700 $ & $ -60.0 $ & $ +12.5 $ & \cite{neese08}  \\
$ $ & $ $ & $ $ & $ $ & $ $ & $ $ & $ $ \\
${\rm BTO-MLC}                       $ \qquad & $ 350 $ & $ 300 $ & $ 434 $ & $ -0.7  $ & $  +0.5 $ & \cite{kar-narayan10,defay13}  \\
$ $ & $ $ & $ $ & $ $ & $ $ & $ $ & $ $ \\
	$                                    $ \qquad & $     $ & $ 110~([111]_{\rm pc}) $ & $     $ & $ -11.0 $ & $ +10.0 $ &    \\
${\rm BCO}                           $ \qquad & $ 300 $ & $     $ & $ 345 $ & $       $ & $       $ & {\rm This~work}  \\
$                                    $ \qquad & $     $ & $ 500~([001]_{\rm pc}) $ & $     $ & $ +60.0 $ & $ -50.0 $ &    \\
$ $ & $ $ & $ $ & $ $ & $ $ & $ $ & $ $ \\
\hline
\hline
\end{tabular}
\label{tab:ecperform}
\caption{Electrocaloric thin film materials. $T$ represents working temperature, $\Delta {\cal E}$ applied electric
	 field (in the BCO case, we also indicate the corresponding orientation in pseudocubic Cartesian notation), 
	 $C_{0}$ heat capacity at zero field, $\Delta S$ isothermal entropy change, and $\Delta T$ adiabatic 
         temperature change. The chemical composition of the materials are: PbZr$_{0.95}$Ti$_{0.05}$O$_{3}$ (PZT),
         poly(vinylidene fluoride-trifluoroethylene)(55/45 mol\%) polymers [P(VDF-TrFE)],
         multilayer capacitor of doped BaTiO$_{3}$ (BTO-MLC), and BiCoO$_{3}$ (BCO).}
\end{table*}

The predicted giant $\Delta T$ and $\Delta S$ values, which can be achieved 
with relatively small driving fields, turn epitaxially grown BCO thin films into very attractive
EC materials. Table~1 offers a comparison with other promising EC
compounds. Note that BCO thin films would operate at $T_{\rm room}$
and display significantly larger adiabatic $T$-spans than observed in
other oxides. Meanwhile, the required electric fields are much smaller than 
reported for FE polymers while similar $\Delta T$ values should be 
attained in both cases. (The smaller heat capacity of perovskite 
oxides contributes favorably to this latter outcome, see Table~1 and Eq.(\ref{eq6}).) 
Hence, BCO films somehow unite the best of the FE ceramics
and FE polymers worlds. On the down side, the reported ${\cal E}$--driven 
BCO phase transitions are strongly discontinuous and thus irreversibility 
issues are likely to appear in practical cooling applications \cite{moya14}.

Let us stress that the coexistence of giant direct and inverse EC effects at room 
temperature offers the possibility of designing improved EC refrigeration cycles
(as originally proposed by Ponomareva and Lisenkov for Ba$_{0.5}$Sr$_{0.5}$TiO$_{3}$
alloys \cite{ponomareva12} and subsequently discussed by Marathe \emph{et al.} in
BaTiO$_{3}$ \cite{marathe17}). In the present case, coexistence of direct and 
inverse EC effects suggests a refrigeration cycle based on the direct transformation 
between the high-entropy FE ${\cal T}$ and low-entropy FE ${\cal M}$ phases as 
induced by ${\cal E}$ rotation, with a cooling performance equal to the 
sum of the individual ${\cal O} \leftrightarrow {\cal T}$ and ${\cal O} \leftrightarrow 
{\cal M}$ cycles. In addition, we foresee alternative strategies for tuning and optimizing 
BCO's EC behavior beyond epitaxial strain. Chemical substitution, for instance, 
represents an obvious option for stabilizing phases that are similar to the ${\cal M}$ 
(alike to the ground state of BiFeO$_{3}$ \cite{cazorla15}) and ${\cal O}$ 
(most common among perovskites \cite{peng18}) polymorphs discussed here, and 
to control the corresponding magnetic transition temperatures. One particular
example is provided by BiCo$_{1-x}$Fe$_{x}$O$_{3}$ solid solutions, where a 
morphotropic transition between a ${\cal T}$--like and a ${\cal M}$--like phase 
is observed to occur at room temperature \cite{azuma08,azuma18}. Likewise, bulk Bi$_{1-x}$La$_{x}$CoO$_{3}$ 
appears to be a good candidate where to realize field-driven ${\cal O} \to {\cal T}$ 
transformations \cite{cazorla17}. Hence BCO offers a variety of experimental 
possibilities to achieve giant EC, bringing new exciting prospects to the field 
of solid-state cooling.

\section{Conclusions}
\label{sec:conclusion}
We have disclosed a number of multiferroic phase transitions in epitaxially grown BCO thin films 
that render giant and dual (that is, direct and inverse) room-temperature EC effects ($|\Delta T| 
\sim 10$~K) driven by moderate electric fields ($\sim 100$~kVcm$^{-1}$). The multiferroic
phase transitions and accompanying EC effects are originated by strong spin-phonon
couplings and polymorph competition, which can be efficiently tuned over a wide 
operating-temperature interval by means of epitaxial strain. Our findings in BCO thin 
films should stimulate the development of cooling devices based on multiferroic EC materials, 
whose energy efficiency and refrigerant performance are good as compared to magnetocaloric and
mechanocaloric materials. Unlike mechanocaloric materials driven by uniaxial stress or 
hydrostatic pressure, scalability down to transistor sizes and implementation on circuitry 
are possible without compromising cooling performance. Unlike magnetocaloric compounds, 
the generation of large external fields at great expense is not necessary. Thus, we 
hope that our results in multiferroic thin films will bring new prospects to the field 
of solid-state cooling.

\section*{Acknowledgments}
This research was supported under the Australian Research Council's Future Fellowship funding
scheme (project number FT140100135) and the Luxembourg National Research Fund through the PEARL
(Grant P12/4853155 COFERMAT) and CORE (Grant C15/MS/10458889 NEWALLS) programs. Computational
resources and technical assistance were provided by the Australian Government and the Government
of Western Australia through Magnus under the National Computational Merit Allocation Scheme and
The Pawsey Supercomputing Centre.

\section*{Appendix: Inclusion of spin-phonon couplings in quasi-harmonic free energy calculations}
\label{sec:appendix}
To calculate the Helmholtz free energy of the competitive polymorphs in $(001)$-oriented 
BCO thin films as a function of temperature ($T$) and in-plane lattice parameter ($a_{\rm in}$), 
we employ a variant of the approaches described in works \cite{cazorla17,cazorla13}.  
Our approach takes into consideration the couplings between magnetic spin disorder and lattice 
phonons, which strongly depend on the temperature and epitaxial strain conditions. Specifically, we 
start by expressing the internal energy of the thin film as:
\begin{eqnarray}
\tilde{E}_{\rm harm}(T, a_{\rm in}) & = & \tilde{E}_{0}(T, a_{\rm in}) + \nonumber \\
 & & \frac{1}{2} \sum_{mn} \tilde{\Xi}_{mn}(T, a_{\rm in}) u_{m} u_{n}~,
\label{eq1app}
\end{eqnarray}
where $\tilde{E}_{0}$ represents an effective static energy, $\tilde{\Xi}_{mn}$ an
effective force-constant matrix, $u$'s atomic displacements, and the dependences of the 
various terms on $T$ and $a_{\rm in}$ are explicitly noted. 

For the ${\cal O}$ and ${\cal M}$ phases ($c \approx a$), the quantities entering Eq.~(\ref{eq1app}) can be calculated 
as:
\begin{eqnarray}
\tilde{E}_{0}^{{\cal O},{\cal M}}(a_{\rm in},T) & = &  E_{0}(a_{\rm in}) + 3 \gamma_{a} (a_{\rm in},T) |S|^{2} J_{a}^{(0)} \, ,\\  
\tilde{\Xi}_{mn}^{{\cal O},{\cal M}}(a_{\rm in},T) & = &  \Phi_{mn}^{0}(a_{\rm in}) + \nonumber \\
                                                      &   &  6 \gamma_{a} (a_{\rm in},T) |S|^{2} J^{(2)}_{a, mn} \, , 
\label{eq2app}
\end{eqnarray}
where $\gamma_{a} (a_{\rm in},T) \equiv \langle S_{i}S_{j} \rangle / |S|^{2}$ represents the 
correlation function between neighboring spins (see Fig.\ref{fig1}d) and $\langle ... \rangle$ the 
thermal average as obtained from our Monte Carlo simulations of the corresponding spin Heisenberg 
hamiltonian (see Sec.~\ref{subsec:MC}). The rest of parameters in $\tilde{E}_{0}^{{\cal O},{\cal M}}$ 
and $\tilde{\Xi}_{mn}^{{\cal O},{\cal M}}$ correspond to:
\begin{eqnarray}
&& E^{0} = \frac{1}{2}\left( E^{\rm FM}_{\rm eq} + E^{\rm G}_{\rm eq}
  \right) \, ,\\
&& \Phi_{mn}^{0} = \frac{1}{2}\left( \Phi_{mn}^{{\rm FM}} +
\Phi_{mn}^{{\rm G}} \right) \, ,\\
&& J_{a}^{(0)} = \frac{1}{6|S|^{2}}\left( E^{\rm FM}_{\rm eq} - E^{\rm
  G}_{\rm eq} \right) \, ,\\
&& J^{(2)}_{a, mn} = \frac{1}{6|S|^{2}}\left( \Phi_{mn}^{{\rm FM}} -
\Phi_{mn}^{{\rm G}} \right) \, .
\label{eq3app}
\end{eqnarray}
In the equations above, superscripts ``FM'' and ``G'' represent perfect
ferromagnetic and antiferromagnetic G-type spin arrangements,
respectively. The $J_{a}^{(0)}$ parameter describes the magnetic
interactions when the atoms remain frozen at their equilibrium
positions; typically, this captures the bulk of the exchange couplings. Meanwhile, the 
$J_{a, mn}^{(2)}$ parameter captures the dependence of the phonon spectrum on the spin
configuration (i.e., spin-phonon coupling effects).

For the ${\cal T}$ phase ($c > a$), we express the corresponding static energy and force constant matrix as:
\begin{eqnarray}
\tilde{E}_{0}^{\cal T}(a_{\rm in},T) & = & E_{0}(a_{\rm in}) +  2 \gamma_{a} (a_{\rm in},T) |S|^{2} J_{a}^{(0)} +  \\ \nonumber
 & & \gamma_{c} (a_{\rm in},T) |S|^{2} J_{c}^{(0)} + \\ \nonumber
 & & 4 \gamma_{ac} (a_{\rm in},T) |S|^{2} J_{ac}^{(0)}  \, ,  
\label{eq4aapp}
\end{eqnarray}
and
\begin{eqnarray}
\tilde{\Xi}_{mn}^{\cal T}(a_{\rm in},T) & = &  \Phi_{mn}^{0}(a_{\rm in}) + \\ \nonumber
	                                &   & 4 \gamma_{a} (a_{\rm in},T) |S|^{2} J^{(2)}_{a,mn} + \\ \nonumber
                                        &   & 2 \gamma_{c} (a_{\rm in},T) |S|^{2} J^{(2)}_{c,mn} + \\ \nonumber
					&   & 8 \gamma_{ac} (a_{\rm in},T) |S|^{2} J^{(2)}_{ac,mn}  \, , 
\label{eq4bapp}
\end{eqnarray}
where $\gamma_{\alpha} (a_{\rm in},T) \equiv \langle S_{i}S_{j} \rangle /
|S|^{2}$, with $\alpha = a, b, ac$, represent the correlation
functions between in-plane and out-of-plane neighboring spins (see Fig.\ref{fig1}d).
(We note that in this case we need to differentiate among in-plane and out-of-plane 
neighboring spins due to the highly anisotropic nature of the tetragonal phase.)
The rest of parameters in $\tilde{E}_{0}^{\cal T}$ and $\tilde{\Xi}_{mn}^{\cal T}$
can be obtained as:
\begin{eqnarray}
&& E^{0} = \frac{1}{4}\left( E^{\rm FM}_{\rm eq} + E^{\rm A}_{\rm eq} + E^{\rm C}_{\rm eq} + E^{\rm G}_{\rm eq}  \right) \, ,\\
&& \Phi_{mn}^{0} = \frac{1}{4}\left( \Phi_{mn}^{{\rm FM}} +  \Phi_{mn}^{{\rm A}} +  \Phi_{mn}^{{\rm C}} + \Phi_{mn}^{{\rm G}} \right) \, ,\\
&& J_{a}^{(0)} = \frac{1}{8|S|^{2}}\left( E^{\rm FM}_{\rm eq} + E^{\rm A}_{\rm eq} - E^{\rm C}_{\rm eq} - E^{\rm G}_{\rm eq} \right) \, ,\\
&& J^{(2)}_{a, mn} = \frac{1}{8|S|^{2}}\left( \Phi_{mn}^{{\rm FM}} + \Phi_{mn}^{{\rm A}} - \Phi_{mn}^{{\rm C}} - \Phi_{mn}^{{\rm G}} \right) \, ,\\
&& J_{c}^{(0)} = \frac{1}{4|S|^{2}}\left( E^{\rm FM}_{\rm eq} - E^{\rm A}_{\rm eq} + E^{\rm C}_{\rm eq} - E^{\rm G}_{\rm eq} \right) \, ,\\
&& J^{(2)}_{c, mn} = \frac{1}{4|S|^{2}}\left( \Phi_{mn}^{{\rm FM}}  - \Phi_{mn}^{{\rm A}} + \Phi_{mn}^{{\rm C}} - \Phi_{mn}^{{\rm G}} \right) \, ,\\
&& J_{ac}^{(0)} = \frac{1}{16|S|^{2}}\left( E^{\rm FM}_{\rm eq} - E^{\rm A}_{\rm eq} - E^{\rm C}_{\rm eq} + E^{\rm G}_{\rm eq} \right) \, ,\\
&& J^{(2)}_{ac, mn} = \frac{1}{16|S|^{2}}\left( \Phi_{mn}^{{\rm FM}} - \Phi_{mn}^{{\rm A}} - \Phi_{mn}^{{\rm C}} + \Phi_{mn}^{{\rm G}}  \right) \, .
\label{eq5app}
\end{eqnarray}
In the equations above, superscripts ``FM'', ``G'', ``A'', and ``C'' mean perfect ferromagnetic, 
antiferromagnetic G-type, antiferromagnetic A-type, and antiferromagnetic C-type spin arrangements, 
respectively. We note that the explained first-principles quasi-harmonic approach could also be employed 
to estimate pyroelectric coefficients \cite{liu16,liu18} in multiferroic materials.


\begin{thebibliography}{30}
\bibitem{pecharsky97} Pecharsky, V. K. $\&$  Gschneidner, Jr. K. A.
                      Giant magnetocaloric effect in Gd$_{5}$(Si$_{2}$Ge$_{2}$). 
                      \textit{Phys. Rev. Lett.} \textbf{78}, 4494 (1997).

\bibitem{bonnot08} Bonnot, E., Romero, R., Ma${\rm \tilde{n}}$osa, Ll., Vives, E. $\&$ Planes, A.
                   Elastocaloric effect associated with the martensitic transition in shape-memory alloys.
                   \textit{Phys. Rev. Lett.} \textbf{100}, 125901 (2008).

\bibitem{cazorla16} Cazorla, C $\&$ Errandonea, D.
                    Giant mechanocaloric effects in fluorite-structured superionic materials.
                    \textit{Nano Lett.} \textbf{16}, 3124 (2016).

\bibitem{cazorla17a} Sagotra, A. K., Errandonea, D. $\&$ Cazorla, C.
                     Mechanocaloric effects in superionic thin films from atomistic simulations.
                     \textit{Nat. Commun.} \textbf{8}, 963 (2017).

\bibitem{cazorla17b} Aznar, A., Lloveras, P., Romanini, M., Barrio, M., Tamarit, J. Ll., Cazorla, C., 
                     Errandonea, D., Mathur, N. D., Planes, A., Moya, X. $\&$  Ma${\rm \tilde{n}}$osa, Ll.
                     Giant barocaloric effects over a wide temperature range in superionic conductor AgI.
                     \textit{Nat. Commun.} \textbf{8}, 1851 (2017).

\bibitem{cazorla18} Sagotra, A. K., Chu, D. $\&$ Cazorla, C.
	            Room-temperature mechanocaloric effects in lithium-based superionic materials.
		    \textit{Nat. Commun.} \textbf{9}, 3337 (2018).

\bibitem{scott11} Scott, J. F.
                  Electrocaloric Materials.
                  \textit{Annu. Rev. Mater. Res.} \textbf{41}, 229 (2011).

\bibitem{kar-narayan10} Kar-Narayan, S. $\&$ Mathur, N. D.
                        Direct and indirect electrocaloric measurements using multilayer capacitors.
                        \textit{J. Phys. D: Appl. Phys.} \textbf{43}, 032002 (2010).  

\bibitem{defay13} Defay, E., Crossley, S., Kar-Narayan, S., Moya, X. $\&$ Mathur, N. D.
                  The Electrocaloric efficiency of ceramic and polymer films.
                  \textit{Adv. Mater.} \textbf{25}, 3337 (2013). 

\bibitem{jiang17} Jiang, Z., Prokhorenko, S., Prosandeev, S., Nahas, Y., Wang, D., ${\rm \acute{I}}$${\rm \tilde{n}}$iguez, J., 
	          Defay, E. $\&$ Bellaiche, L. 
		  Electrocaloric effects in the lead-free Ba(Zr,Ti)O$_{3}$ relaxor ferroelectric from atomistic simulations.
		  \textit{Phys. Rev. B} \textbf{96}, 014114 (2017).

\bibitem{geng15} Geng, W., Liu, Y., Meng, X., Bellaiche, L., Scott, J. F., Dkhil, B. $\&$ Jiang, A.
	         Giant negative electrocaloric effect in antiferroelectric La‐doped Pb(Zr,Ti)O$_{3}$ thin films near room temperature.
		 \textit{Adv. Mater.} \textbf{27}, 3165 (2015).

\bibitem{mangeri16} Mangeri, J., Pitike, K. C., Alpay, S. P. $\&$ Nakhmanson, S.
	            Amplitudon and phason modes of electrocaloric energy interconversion.
		    \textit{NPJ Comput. Mater.} \textbf{2}, 16020 (2016).

\bibitem{zhang11} Zhang, J., Alpay, S. P. $\&$ Rossetti, J. A.
	          Influence of thermal stresses on the electrocaloric properties of ferroelectric films.
		  \textit{Appl. Phys. Lett.} \textbf{98}, 132907 (2011).

\bibitem{tong14} Tong, T., Karthik, J., Mangalam, R. V. K., Martin, L. W. $\&$ Cahill, D. G.
	         Reduction of the electrocaloric entropy change of ferroelectric PbZr$_{1−x}$Ti$_{x}$O$_{3}$ epitaxial 
		 layers due to an elastocaloric effect.
		 \textit{Phys. Rev. B} \textbf{90}, 094116 (2014).

\bibitem{pandya17} Pandya, S., Wilbur, J. D., Bhatia, B., Damodaran, A. R., Monachon, C., Dasgupta, A., King, W. P., 
	           Dames, C. $\&$ Martin, L. W. 
		   Direct measurement of pyroelectric and electrocaloric effects in thin films.  
		   \textit{Phys. Rev. Appl.} \textbf{7}, 034025 (2017). 

\bibitem{defay18} Defay, E., Faye, R., Despesse, G., Strozyk, H., Sette, D., Crossley, S., Moya, X. $\&$ Mathur, N. D.
                  Enhanced electrocaloric efficiency via energy recovery.
                  \textit{Nat. Commun.} \textbf{9}, 1827 (2018).

\bibitem{mischenko06} Mischenko, A. S., Zhang, Q., Scott, J. F., Whatmore, R. W. $\&$ Mathur, N. D.
                      Giant electrocaloric effect in thin-film PbZr$_{0.95}$Ti$_{0.05}$O$_{3}$.
                      \textit{Science} \textbf{311}, 1270 (2006).  

\bibitem{neese08} Neese, B., Chu. B., Lu. S.-G., Wang, Y., Furman, E. $\&$ Zhang, Q. M.
                  Large electrocaloric effect in ferroelectric polymers near room temperature.
                  \textit{Science} \textbf{321}, 821 (2008).

\bibitem{asbani17}  Asbani, B., Gagou, Y., Dellis, J.-L.,  Trcek, M., Kutnjak, Z., Amjoud, M., Lahmar, A., Mezzane, D. 
	            $\&$ El Marssi, M.
		    Lead free Ba$_{0.8}$Ca$_{0.2}$Te$_{x}$Ti$_{1-x}$O$_{3}$ ferroelectric ceramics exhibiting high 
		    electrocaloric properties.
		    \textit{J. Appl. Phys.} \textbf{121}, 064103 (2017).

\bibitem{cazorla17} Cazorla, C., Di\'eguez, O. $\&$ ${\rm \acute{I}}$${\rm \tilde{n}}$iguez, J.
                    Multiple structural transitions driven by spin-phonon couplings in a perovskite oxide.
                    \textit{Sci. Adv.} \textbf{3}, e1700288 (2017).

\bibitem{cazorla13} Cazorla, C. $\&$ ${\rm \acute{I}}$${\rm \tilde{n}}$iguez, J.
                    Insights into the phase diagram of bismuth ferrite from quasiharmonic free-energy calculations.
                    \textit{Phys. Rev. B} \textbf{88}, 214430 (2013).

\bibitem{pbe96} Perdew, J. P., Burke, K. $\&$ Ernzerhof, M.
                Generalized gradient approximation made simple.
                \textit{Phys. Rev. Lett.} \textbf{77}, 3865 (1996).

\bibitem{vasp} Kresse, G. $\&$ F\"urthmuller, J.
               Efficient iterative schemes for ab initio total-energy calculations using a plane-wave basis set.
               \textit{Phys. Rev. B} \textbf{54}, 11169 (1996);
               Kresse, G. $\&$ Joubert, D.
               From ultrasoft pseudopotentials to the projector augmented-wave method.
               \textit{Phys. Rev. B} \textbf{59}, 1758 (1999).

\bibitem{bloch94} Bl\"ochl P. E.
                  Projector augmented-wave method.
                  \textit{Phys. Rev. B} \textbf{50}, 17953 (1994).

\bibitem{cazorla15} Cazorla, C. $\&$ Stengel, M.
	            Electrostatic engineering of strained ferroelectric perovskites from first-principles.
		    \textit{Phys. Rev. B} \textbf{92}, 214108 (2015). 

\bibitem{cazorla17c} Cazorla, C.
                     Lattice effects on the formation of oxygen vacancies in perovskite thin films.
                     \textit{Phys. Rev. Appl.} \textbf{7}, 044025 (2017).

\bibitem{kresse95} Kresse, G., Furthm\"uller, J. $\&$ Hafner, J.
                   Ab initio force constant approach to phonon dispersion relations of diamond and graphite.
                   \textit{Europhys. Lett.} \textbf{32}, 729 (1995).

\bibitem{alfe09} Alf\`e, D.
                 PHON: a program to calculate phonons using the small displacement method.
                 \textit{Comp. Phys. Commun.} \textbf{180}, 2622 (2009).

\bibitem{cazorla17d} Cazorla, C. $\&$ Boronat, J.
	             Simulation and understanding of atomic and molecular quantum crystals.
		     \textit{Rev. Mod. Phys.} \textbf{89}, 035003 (2017).
	 
\bibitem{strecka15} Stre\v{c}ka, J. $\&$ Ja\v{s}\v{c}ur, M.
                    A brief account of the Ising and Ising-like models:
                    mean-field, effective-field and exact results.
                    \textit{Acta Phys. Slovaca} \textbf{65}, 235-367 (2015).

\bibitem{moya14} Moya, X., Kar-Narayan, S. $\&$ Mathur, N. D.
                 Caloric materials near ferroic phase transitions.
                 \textit{Nat. Mater.} \textbf{13}, 439-450 (2014).

\bibitem{manosa17} Ma${\rm \tilde{n}}$osa, L. $\&$ Planes, A.
                   Materials with giant mechanocaloric effects: Cooling by strength.
                   \textit{Adv. Mater.} \textbf{29}, 1603607 (2017).

\bibitem{belik06} Belik, A. A., Iikubo, S., Kodama, K., Igawa, N., Shamoto, S., Niitaka, S., Azuma, M., Shimakawa, Y.,
         Takano, M., Izumi, F. $\&$ Takayama-Muromachi, E.
          Neutron powder diffraction study on the crystal and magnetic structures of BiCoO$_{3}$.
          \textit{Chem. Mater.} \textbf{18}, 798-803 (2006).

\bibitem{king93} King-Smith, R. D. $\&$ Vanderbilt, D.
	         Theory of polarization of crystalline solids.
		 \textit{Phys. Rev. B} \textbf{47}, 1651(R) (1993). 
  
  
\bibitem{oka10} Oka, K., Azuma, M., Chen, W.-T., Yusa, H., Belik, A. A., Takayama-Muromachi, E., Mizumaki, M., Ishimatsu, N.,
                Hiraoka, N., Tsujimoto, M., Tucker, M. G., Attfield, J. P. $\&$ Shimakawa, Y.
                Pressure-induced spin-state transition in BiCoO$_{3}$.
                \textit{J. Am. Chem. Soc.} \textbf{132}, 9438 (2010).

\bibitem{ponomareva12} Ponomareva, I. $\&$ Lisenkov, S.
	          Bridging the macroscopic and atomistic descriptions of the electrocaloric effect. 
                  \textit{Phys. Rev. Lett.} \textbf{108}, 167604 (2012).

\bibitem{marathe17} Marathe, M., Renggli, D., Sanlialp, M., Karabasov, M. O., Shvartsman, V. V., Lupascu, D. C., 
	           Gr\"unebohm, A. $\&$ Ederer, C.
		   Electrocaloric effect in BaTiO$_{3}$ at all three ferroelectric transitions:
		   Anisotropy and inverse caloric effects.
		   \textit{Phys. Rev. B} \textbf{96}, 014102 (2017).

\bibitem{peng18} Chen, P., Grisolia, M. N., Zhao, H. J., González-Vázquez, O. E., Bellaiche, L., Bibes, M., Liu, B.-G. $\&$
                 ${\rm \acute{I}}$${\rm \tilde{n}}$iguez, J.
                 Energetics of oxygen-octahedra rotations in perovskite oxides from first principles.
                 \textit{Phys. Rev. B} \textbf{97}, 024113 (2018).

\bibitem{azuma08} Azuma, M., Niitaka, S., Hayashi, N., Oka, K., Takano, M., Funakubo, H. $\&$ Shimakawa, Y.
              Rhombohedral-tetragonal phase boundary with high Curie temperature in $(1-x)$BiCoO$_{3}$-$x$BiFeO$_{3}$ solid
              solution. \textit{Jpn. J. Appl. Phys.} \textbf{47}, 7579 (2008).

\bibitem{azuma18} Hojo, H., Oka, K., Shimizu, K., Yamamoto, H., Kawabe, R. $\&$ Azuma, M.
                  Development of bismuth ferrite as a piezoelectric and multiferroic material by cobalt substitution.
                  \textit{Adv. Mater.}, 1705665 (2018).

\bibitem{liu16} Liu, J., Fern\'andez-Serra, M. V. $\&$ Allen, P. B. 
	        First-principles study of pyroelectricity in GaN and ZnO.
		\textit{Phys. Rev. B} \textbf{93}, 081205(R) (2016).

\bibitem{liu18} Liu, J. $\&$ Pantelides, S. T.
	        Mechanisms of pyroelectricity in three- and two-dimensional materials.
		\textit{Phys. Rev. Lett.} \textbf{120}, 207602 (2018). 

\end{thebibliography}
\end{document}